\documentstyle[a4,12pt,epsf]{article}
\begin{document}
\def\abstract#1{\begin{center}{\large ABSTRACT}\end{center}
\par #1}
\def\title#1{\begin{center}{\LARGE #1}\end{center}}
\def\author#1{\begin{center}{\sc #1}\end{center}}
\def\address#1{\begin{center}{\it #1}\end{center}}
\def\pubnum{302/COSMO-59 }

\hfill
\parbox{6cm}{{TIT/HEP-\pubnum} \par October, 1995 }
\par
\vspace{7mm}
\title{Causality violation and singularities}
\vskip 1cm
\author{ Kengo Maeda \footnote{e-mail:maeda@th.phys.titech.ac.jp}
and Akihiro Ishibashi 
\footnote{e-mail:akihiro@th.phys.titech.ac.jp}}
\address{Department of Physics, Tokyo Institute of \\ Technology,
Oh-Okayama Meguro-ku, Tokyo 152, Japan}
\vskip 1 cm
\abstract{We show that singularities necessarily
occur when a boundary of causality violating set exists in a space-time 
under the physically suitable assumptions except the global
causality condition in the Hawking-Penrose
singularity theorems. Instead of the global causality condition, 
we impose some restrictions on the causality violating sets to
show the occurrence of singularities.}

\section{Introduction}

Space-time singularities have been discussed for a long time 
in general relativity. 
In 1970, Hawking and Penrose[1] showed that 
singularities, which mean causal geodesic 
incompleteness, could occur in a space-time 
under seemingly reasonable conditions 
in classical gravity.
Their singularity theorem has an important implication that
our universe has an initial singularity if we do not consider
quantum effects. However, this theorem is physically unsatisfactory
in the sense that the causality requirement everywhere 
in a space-time
seems too restrictive. We can only experience local events and 
there is no guarantee that the causality holds in the
entire universe. As is well known, Kerr type black holes  
have causality violating sets if the space-time is
maximally extended. 
Therefore, it will be important to investigate occurrence of 
singularities in a space-time in which the global causality condition 
is violated.

There are some works on a causality violation concerned with
the occurrence of singularity.
Tipler[2,3] showed that any attempt to evolve closed
timelike curves from an initial regular Cauchy data would cause 
singularities to form in a space-time. 
He presented a singularity theorem in which the global 
causality condition in the Hawking-Penrose theorem 
is replaced by the weaker one and 
adding the stronger energy condition.
In his theorem his stronger energy condition is essential to
the occurrence of singularities. 

Kriele presented his singularity theorems 
in which causality violating sets 
are restricted but with usual energy condition 
in the Hawking-Penrose theorem 
instead of Tipler's energy condition.
He showed that the causality violating set has incomplete null 
geodesics if its boundary is compact~[4]. 
Kriele[5] also presented a generalization of
the Hawking-Penrose singularity theorem. In his paper he 
showed that singularities would occur provided that causality 
holds at least in the future endpoints of the trapped set.
 
Newman[6] found a black hole solution which had no 
singularities. 
This black hole solution is obtained by a suitable conformal 
transformation of the G\"{o}del universe: One might consider 
his conclusion would suggest that causality violating set can 
prevent singularities from occurring. However, his case 
seems too special and even unphysical, 
because causality is violated in the entire space-time. 
It is physically more acceptable to assume 
that at least there must be causality
preserving regions in a space-time. 
One can pick up the Taub-NUT universe as an example which
contains both causality violating and preserving sets. 
In this universe, there exist singularities 
on the boundary of causality violating
sets. 
This suggests that the boundary generates a 
geodesic incompleteness. 

In this paper we shall show that the boundary of causality 
violating sets are essential to occurrence of
singularities. We also discuss relation between our theorems 
and the Hawking-Penrose theorem. 

In the next section, we briefly review Tipler's and Kriele's
singularity theorems. In section 3, the definitions 
and the lemmas for discussing causal structure and singularities 
are listed up. We present our singularity theorems for partially 
causality violating space-times in section 4. 
Section 5 is devoted to summary. 

\section{Tipler's and Kriele's theorems}

We review Tipler's and Kriele's theorems in this section.
In addition, we discuss how causality violation is 
related to singularities in these theorems. 

First, we quote Tipler's theorem. \\
\\
Tipler's theorem(1977)\\
A space-time $(M,g)$ cannot be null geodesically complete if\\
(1) $R_{ab}K^{a}K^{b}\ge 0$ for all null vectors $K^{a}$;\\
(2) there is a closed trapped surface in $M$;\\
(3) the space-time is asymptotically deterministic, and the Einstein
equations hold;\\
(4) the partial Cauchy surface defined by (3) is non-compact.\\ 
\\
Here the asymptotically deterministic condition in the
condition (3) is defined as follows. \\
\\
{\it Definition}\\
A space-time $(M,g)$ is said to be {\it asymptotically deterministic}
if\\
(i) $(M,g)$ contains a partial Cauchy surface $S$ such that\\
(ii) either $H(S)=H^{+}(S)\cup H^{-}(S)$ is empty, or if not, then 
\[\lim_{s \to a} [\inf T_{ab}K^{a}K^{b}] >0 \]
on at least one of the null geodesic generators ${\gamma}(s)$ of
$H(S)$, where $a$ is the past limit of the affine parameter along 
$\gamma$ if $\gamma\in H^{+}(S)$, and the future limit if $\gamma\in
H^{-}(S)$.
($K^{a}$ is the tangent vector to $\gamma$.)\\
\\
This condition has been introduced by following reasons. 
In the case that the formation of a Cauchy horizon $H^+(S)$ 
is due to causality violation, one would expect that 
the region where $H^+(S)$ begins 
would contain enough matter (the condition (ii)) 
which causes gravitational field sufficiently strong so as to 
tip over the light cones and eventually leads to causality 
violation. We can regard this condition as a special type of energy
condition which dispenses with the causality condition 
in the Hawking-Penrose singularity theorem. 

In the following sections, we shall impose some conditions on 
causality violating sets to replace global 
causality condition in the Hawking-Penrose theorem 
instead of imposing this energy condition.  

Next, we quote some definitions and Kriele's theorems~[4, 5]. \\
\\
{\it Definition}\\
${\bullet}${\it  focal point}\\  
Let $S$ be a locally spacelike surface ( not necessary achronal
surface)
and let us consider a future directed null geodesic, ${\beta}(t)$, 
from $S$ parameterized by
$t$. If for any point ${\beta}(t)$ such that $ t \ge t_{1}$, there 
exists an arbitrarily close timelike curve from $S$ to the point 
${\beta}(t)$, then ${\beta}$ is called a {\it focal point} to $S$.
\\ \\
${\bullet}${\it Generalized future horismos of $S$}\\ 
{\it Generalized future horismos of} $S$, denoted by
$e^{+}(S,M)$, is a
closure of all future null geodesics $\beta$ from $S$
which have no focal points. (The future end points of $e^{+}(S,M)$
correspond 
to the focal points.) \\  
\\
${\bullet}${\it cut locus}: $cl(S,M,+)$\\  
The set of future end points of $e^{+}(S,M)$.\\
\\
${\bullet}${\it  almost closed causal curve}\\
Choose an arbitrary Riemannian metric $h$ of $M$. Let $\alpha$ be a
curve and $\beta$
be a reparametrization of $\alpha$ with
$h({\beta}',{\beta}')=1$. 
Then $\alpha$ is called {\it almost closed} 
if there exists an $X\in {\beta}'(t)$
such that for every neighbourhood $U$ of $X$ in the tangent
bundle, $TM$, there exists a deformation 
$\gamma$ of $\beta$ in ${\pi}_{TM}(U)$ which 
yields a closed curve and satisfies 
${\gamma}(t)\in{\pi}(U)\Rightarrow {\gamma}'(t)\in U$.\\
\\
Kriele's theorem\\
Theorem 1(1990)\\
$(M,g)$ is causal geodesically incomplete if:\\ 
(1) $R_{ab}K^aK^b \ge 0$ for every causal vector $K^a$ 
and the generic
condition is satisfied.\\
(2) (a) there exists a closed locally spacelike 
but not necessarily achronal trapped surface 
$S$ or (b) there exists a point $r$ such that on every past 
(or every future) null geodesic from $r$ 
the divergence $\theta$ of the null geodesics 
from $r$ becomes negative or (c) there exists 
a compact achronal set $S$ without edge.\\
(3) neither $cl(S,M,+)$ (respectively $cl(r,M,+)$) nor 
any $cl(D,M,-)$, where $D$ is a compact topological 
submanifold (possibly with boundary) with $D\cap S \neq
\emptyset$ (respectively $r \in D$), contains any 
almost closed causal curve that is a limit curve of 
a sequence of closed timelike curves.\\

This theorem is the maximum generalization of 
the Hawking-Penrose theorem in the sense that causality 
may be violated 
in the almost all regions except the cut locus.
In this theorem causality violation seems to play a role of 
keeping the space-time under consideration from having 
singularities.    

In theorem 2 below it is shown that there exist singularities when the
causality violating set is compact even if there is no trapped surface.
However, one cannot see which causes singularities, 
the compactness of the causality 
violating set or causality violation itself.\\
\\
Theorem 2(1989)\\
Let $(M,g)$ be a space-time with chronology violating set $V$ that
satisfies \\
(1) $R_{ab}K^aK^b \ge 0$ for every null vector $K^a$ and the generic
condition is satisfied.\\
(2) $V$ has a compact closure but $M-V\neq\, \emptyset$. \\
Then $V$ is empty or $\dot{V}$ is generated by almost closed but
incomplete
null geodesics.\\
\\
\\

\section{Preliminaries}

We consider a space-time $(M,g)$, where $M$ is a 
four-dimensional connected differentiable manifold and $g$ is a 
Lorentzian and suitably differentiable metric.
In this section, we quote some definitions and useful lemmas
from (HE)[1] for the discussion of causal structure and space-time
singularities.\\
\\
{\it Definition} (HE)\\
A point $p$ is said to be a {\it limit point} of an
infinite sequence of non-spacelike curves $l_n$ if every 
neighbourfood of $p$ intersects an infinite number of the
$l_n $'s.\\
A non-spacelike curve $l$ is said to be a {\it limit
curve} of the sequence $l_n$ if there is a subsequence
$l'_n$ of the $\l_n$ such that for every $p \in l$,
$l'_n$ converges to $p$. \\
\\
{\it Proposition 1} (HE 6.4.1)\\
The chronology violating set $V$ of $M$ is the disjoint union of the 
form $I^+(q) \cap I^-(q),q \in M$.\\
\\
{\it Lemma 1} (HE 6.2.1)\\
Let $O$ be an open set and let $l_n$ be an infinite sequence
of non-spacelike curves in $O$ which are future-inextendible in 
$O$. If $p\in O$ is a limit point of $l_n$, then through $p$ 
there is a non-spacelike curve $l$ which is
future-inextendible in $O$ and which is a limit curve of the
$l_n$. \\
\\
{\it Proposition 2} (HE 4.5.10)\\
If $p$ and $q$ are joined by a non-spacelike curve $l(v)$ 
which is not a null geodesic they can also be joined by a 
timelike curve.\\
\\
{\it Proposition 3} (HE 4.4.5) \\
If $R_{ab}K^a K^b \ge 0$ everywhere and if at $p=\gamma(v_1), 
K^c K^d K_{[a}R_{b]cd[e}K_{f]}$ is non-zero, there will be $v_0$ 
and $v_2$ such that $q=\gamma(v_0)$ and $r=\gamma(v_2)$ will be 
conjugate along  $\gamma(v)$ provided $\gamma(v)$ can be
extended to these values.\\
\\
{\it Proposition 4} (HE 4.5.12)\\
If there is a point $r$ in $(q,p)$ conjugate to $q$ along
$\gamma(t)$ then there will be a variation of $\gamma(t)$ which
will give a timelike curve from $q$ to $p$. \\ 
\\
{\it Proposition 5} (HE 6.4.6)\\
If $M$ is null geodesically complete, every inextendible null geodesic
curve has a pair of conjugate points, and chronology condition holds on
$M$, then the strong causality condition holds on $M$. \\
\\
{\it Proposition 6} (HE 6.4.7)\\
If the strong causality condition holds on a compact set $\varphi$,
there
can be no past-inextendible non-spacelike curve totally or partially 
past imprisoned in $\varphi$.\\
\\
{\it Prop.5} physically means that the chronology condition is equivalent
to
the strong causality condition if energy conditions are satisfied. 
\section{The theorem}


Generally, one can consider either of the following two cases 
in which causality violating sets and their boundaries exist. 
One is that there are {\it closed} null geodesic curves 
lying on the boundary or closed non-spacelike curves 
which pass through at
least one point on the boundary. 
\footnote{The case that whole 
null generators of the boundary are closed or imprisoned is 
similar to the situation which Hawking considered~[7]. 
When he discussed the chronology
violating sets appearing in a bounded region of general space-time 
without curvature singularities, he introduced the notion of 
{\it the compactly generated Cauchy horizon} defined as 
a Cauchy horizon such that all the past 
directed null geodesic generators enter and remain
within a compact set. This is analogous to the existence of closed
null curves on the boundary of $V$. He asserted that one cannot make
such a
Cauchy horizon while the weak energy condition is satisfied. 
This also supports our claims.}  
The other is that there is no {\it closed} non-spacelike curve 
which passes through a point on the boundary.    
Here we have used the word 
{\it closed curve} in a specific sense that the curve is closed 
and moreover one lap length of 
the curve does not diverge. 

We will show in each case that such a space-time has
singularities in what follows.\\
\\
{\it Theorem 1 }\\ 
If a space-time $(M,g)$ is null geodesically complete, then the 
following three conditions cannot be all satisfied together: \\
(a) There exists a chronology violating region $V$ which 
 does not coincide with the whole space-time, i.e. $M-V \neq
 \emptyset$, \\
(b) every inextendible non-spacelike geodesic in $(M,g)$ contains a 
pair of conjugate points,\\
(c) there exists at least one point $p$ on the boundary of $V$ such
that
 each closed timelike curve through a point in the 
$V \cap \varepsilon$ can be entirely contained 
in some compact set $K$. 
($\varepsilon$ is an arbitrary small neighbourhood of $p$.) \\    

As mentioned above, if the condition (c) is satisfied, roughly 
speaking, one can always pick out an infinite sequence such that 
the one lap length of each closed timelike curve does not
diverge and their shape does not change abruptly 
when a point on each closed curve approaches to 
the boundary of $V$. 

This condition (c) is satisfied, for example, on the causality
violating sets which cause {\it compactly generated Cauchy 
horizons }~[7].
Causality violating sets of the Taub-NUT universe 
also satisfy the condition (c) because whose boundaries contain
closed null geodesics. Therefore we can apply {\it Theorem 1} to 
the Taub-NUT universe, which indeed has singularities. 

This condition (c) does not require that the boundary $\dot V$ 
is compact. Thus {\it Theorem 1} is essentially different from 
the Kriele's theorem 2.\\ 
\\
{\it Proof}.\\
The chronology violating set $V$ is an open set by {\it Prop 1}.
If $V \neq \emptyset$, from the condition (a), we can find a 
boundary set $\dot V$ in $M-V$. 
Let us consider a sequence of points 
$q_n$ in $V\cap\varepsilon$ which converges to 
$p$ ($\lim_{n \to \infty} q_{n}=p$).
By the definition of $V$ there is 
a closed timelike curve $l_{n}$ through
$q_n$. From the condition (c), there exists a compact set $K$  
such that each $l_{n}$ is entirely contained in $K\cap V$.
Let $l$ be a limit curve of the sequence $l_n$ which passes through
the limit point $p$. Choosing a suitable parameter of each $l_n$
so that $l_n$ is inextendible, the limit curve $l$ 
is also non-spacelike inextendible curve in $K\cap \bar{V}$ by
{\it Lemma 1}.

Let us consider the case 
that the limit point $p\in\dot{J}^{+}(q),\, q
\in V$ without loss of generality.
This limit curve must also be contained in $K\cap \bar{V}$ 
because of the condition (c). 
Therefore $l$ is totally past and future imprisoned 
in $K\cap \bar{V}$. 
If some point $p'$ of $l$ which is in the past of $p$ 
is contained in $V$, there exists a closed 
non-spacelike curve but not
null geodesic through $p$. Because one can connect 
the limit point $p$ to some point $c$ in $V$ in the future of 
the $p$ with some non-spacelike curve $\lambda$, 
one can always find a closed non-spacelike curve but
not a null geodesic one such that $p \rightarrow c 
\rightarrow q \rightarrow p' \rightarrow p$ as depicted in Figure 1. 
This curve can be varied to a closed timelike
curve through $p$ by {\it Prop.2}. This contradicts with the 
achronality of the boundary $\dot V$ in which $p$ is
contained. Therefore 
any point of $l$ in the past of $p$ is not contained in $V$, 
but in the compact set $\dot{J}^{+}(q) \cap K$.
If the null geodesic generator $l$ of $\dot{J}^{+}(q)$ through
$p$ is closed, this generator has no future and past end points.
Then $l$ has pair conjugate points from {\it Prop.3} if $l$ is
complete. This contradicts with the achronality of the boundary
$\dot{V}$. Therefore this null geodesic generator $l$ is not
closed but past imprisoned in the compact set 
$\dot{J}^{+}(q) \cap K$.
Let $p_n \in \{K \cap (M- \bar V)\} $ be an infinite sequence 
which converges to $p\in l$ and $r_n\in \{K \cap (M- \bar V)\}$
be another infinite sequence such that $r_n \in
\dot{J}^{-}(p_n)$ and converges to 
the point $r(\neq p)$ on $l$. 
Then one can take an infinite sequence of curves $\lambda_n$ 
such that each of which is an inextendible null geodesic 
through $p_n$ and $r_n$.  
If $M$ is null geodesically complete, each $\lambda_n$ can
be extended into the open region 
$\{ \dot{J}^-(p_n) \cap I^{-}(K) \}$
because each $\lambda_n$ is entirely contained 
in $M-\bar V$ where 
the strong causality condition holds by using {\it Prop.~5}. 
Therefore, the limit curve $\lambda$ of $\lambda_n$, 
which is an inextendible null geodesic curve through $p$ and $r$ 
from {\it Lemma 1}, is not imprisoned in the compact set 
$\{K \cap \dot{J}^+(q)\} \subset \{K \cap (M - V)\}$. 
However, this contradicts the fact that $\lambda$ 
coincides with $l$ by reparametrization of affine parameter 
since both of them are null geodesics through 
the two points $p$ and $r$. Otherwise, there exists 
a null curve broken at $p$ and $r$ 
which is lying on $\dot{V}$, and it can be 
deformed to a timelike curve. 
This contradicts the achronality of $\dot{V}$.  
\qquad $\Box$\\ 

Combining {\it Theorem 1} and the Hawking-Penrose theorem~[1], 
we immediately get the following corollary. \\
\\
{\it Corollary}\\
If a space-time $(M,g)$ is causally complete, then the following
conditions cannot all hold:\\
(1) every inextendible non-spacelike geodesic contains a pair of 
 conjugate points, \\
(2) the chronology condition holds everywhere on $(M, g)$ or
even if chronology condition is violated somewhere, such a region 
satisfies the condition (c),\\
(3) there exists a future-(or past-)trapped set $S$. \\

We have considered the case that a chronology violating 
set satisfies the condition (c). 
However, the causality violating sets in the Kerr black hole 
do not satisfy the condition (c). So we cannot 
apply our {\it Theorem 1} to the Kerr solution. 
However, we could still prove 
the existence of singularities 
if a given space-time satisfies the condition below. \\
\\
{\it Condition} ($c'$) \\
Let each chronology violating set be $V_{i}$. 
Any $V_i$ is causally separated from $V_{j \neq i}$, i.e. 
$( \dot J^+(q) \cup \dot J^-(q)) \cap V_{j \neq i} 
= \emptyset$ for all $ q\in V_i $. \\ 

For a space-time $(M,g)$ which satisfies 
this condition ($c'$) but not the condition (c), 
we can apply Kriele's theorem 1, taking a set $S$ 
in his theorem~2 as $\dot{J}^{+}(q)\cap \dot{J}^{-}(q)$. 
In usual, we expect that 
the set $\dot{J}^{+}(q)\cap \dot{J}^{-}(q)$ 
is compact, which may have the topology $S^{2}$. 
However, there is a case that 
$\dot{J}^{+}(q)\cap \dot{J}^{-}(q)$ has non-compact topology. 
For example, in the case that $\dot{J}^{+}(q)\cap
\dot{J}^{-}(q)$ has topology $S^1 \times R$, 
we can regard the quotient space 
$e^{+}(\dot{J}^{+}(q)\cap \dot{J}^{-}(q)) / R$ as the
$e^{+}(S)$ in the Kriele's theorem 1, because 
the relevant thing in his theorem is that $e^{+}(S)$ is compact. 

We obtain the following theorem. \\ 
\\
{\it Theorem 2}\\
A space-time $(M,g)$ which satisfies the conditions (a), (b), and 
either (c) or ($c'$) is null or timelike geodesically
incomplete.\\
 
As easily verified from the Penrose diagram of the Kerr
solution, the condition ($c'$) is satisfied for the
Kerr solution. 
This theorem is applicable to the Kerr solution which indeed has 
singularities.  
\\ 
\\ 
{\it proof}. \\
We suppose that $(M,g)$ is null or timelike geodesically complete.
We only have to prove the case that the condition ($c'$) hold 
but the condition (c) does not. In such a space-time $(M,g)$, 
every null geodesic generator on $\dot{V}$ is not closed.  
 
Now we consider a non-closed null geodesic on $\dot{V}$. 
This null geodesic belongs to $\dot{J}^{+} 
(q)$ or $\dot{J}^{-}(q)$ , as $V$ is 
$I^{+}(q)\cap I^{-}(q)$ ($q\in V$). 
Let this null geodesic belong to $\dot{J}^{+}(q)$ without losing 
generality. If this null geodesic has a past end point, it must
be $q$. Let us take a point $p(\neq q)$ on this null geodesic
and also let it be on the $\dot V$. 
Because $q\in V$, there is a closed timelike curve through $q$. 
This means that a timelike curve from $q$
to $p$ exists by {\it Prop.2}. Therefore, $p$ belongs to
$I^{+}(q)$. This
contradicts $p \in \dot{V}$. If this null geodesic has no past end
point,
it is inextendible in the past.   
If the boundary of $V$ contains this null geodesic entirely, from
the condition (b), this boundary can be connected by timelike 
curves by {\it Prop.4}. This is also contradiction to the 
achronality of $\dot V$. 
Hence, let us consider the case that the boundary of $V$ does not
contain 
the whole segment of this null geodesic, that is, the null
geodesic has an end point on the compact surface 
$ S:= \dot J^+(q)\cap \dot J^-(q)$. 
Extending this null geodesic beyond the future end
point, we obtain an inextendible null geodesic lying on
$\dot{J^{+}}(q)$ and call it outgoing. We also obtain an
inextendible null geodesic belongs to $\dot{J^{-}}(q)$ and call it
ingoing. The outgoing null geodesic has a pair of conjugate 
points from the condition (b). One of the conjugate points 
is on the segment lying on the $\dot V$. 
The other is on the segment 
lying on the $ \dot J^+(q)-\dot V$. The ingoing null geodesic
on the $\dot J^-(q)$ also has a pair of conjugate points in the 
same way as the outgoing case. Thus, $S$ plays the same 
role as the trapped surface in the Kriele's theorem 1. 
From the condition ($c'$), the condition (3) of 
Kriele's theorem 1 is satisfied 
in the cut locuses of $S$, intersections of outgoing 
and ingoing null geodesics, 
because the condition (c) is not satisfied
(if the condition (c) is satisfied, there exists an almost closed
causal curve.).
Therefore we can show the existence of singularities from Kriele's
theorem 1. \qquad $\Box$\\ 

\section{Conclusions and discussion}
We have shown that {\it the boundaries of causality 
preserving and violating regions cause singularities 
in a physical space-time}.

We would like to emphasize that 
{\it Theorem 1} supplements the Hawking-Penrose theorem in the sense
that the global causality condition is relaxed 
to some degree and instead, 
the condition (c) or the condition ($c'$) is 
imposed on chronology violating sets. Roughly speaking, it is
possible for observers to talk about the existence 
of singularities assuming that 
{\it our space-time has a causality 
preserving region}, which conforms to our 
experience.

Whether the quasi-global condition ($c'$) is removable or not 
is still an open question.  
 
As well as the Hawking-Penrose theorem, 
our theorem cannot predict where 
singularities exist and how strong they are, which are 
left for future investigations. \\

\begin{center} {\bf \large Acknowledgements} \end{center}

We would like to thank Prof. A.Hosoya and S.Ding 
for useful discussions and helpful suggestions. 
We are also grateful to Prof. H.Kodama for kind advice and 
critical comments.

\clearpage
\begin{figure}[htbp]
 \centerline{\epsfxsize=10.0cm \epsfbox{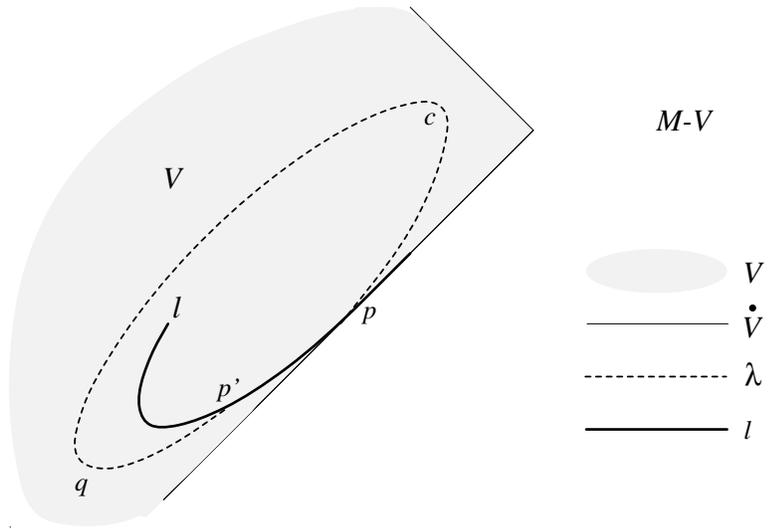}}
 \caption{In the case that the past points of 
the limit curve $l$ go into the $V$, we can find a closed
non-spacelike non-geodesic curve like a $p \rightarrow c 
\rightarrow q \rightarrow p' \rightarrow p$, 
which is the union of $\lambda$ and a segment $p'\rightarrow
p$. }
        \protect 
\end{figure}

\begin{figure}[htbp]
 \centerline{\epsfxsize=12.0cm \epsfbox{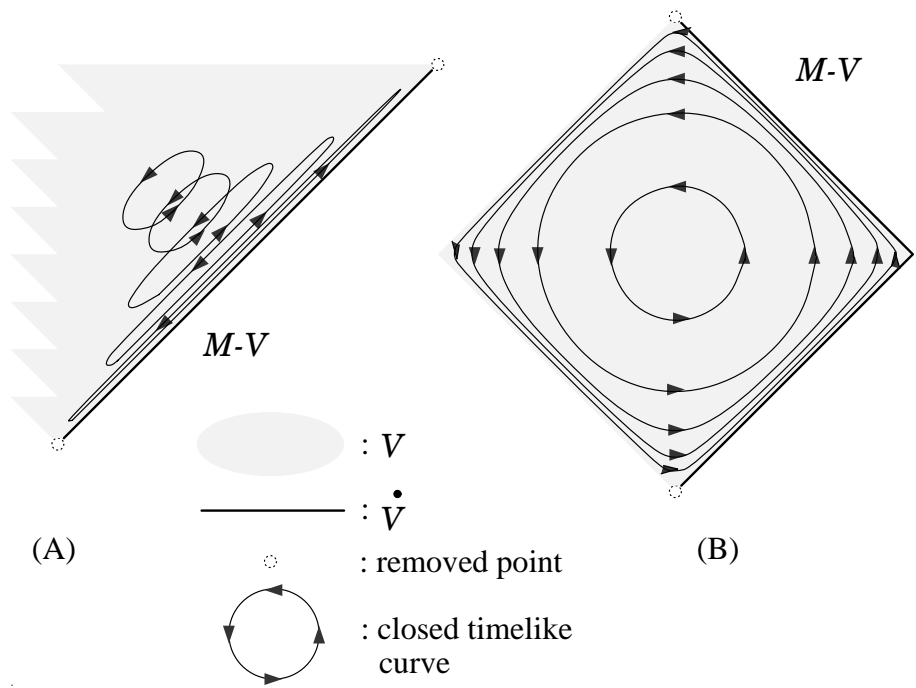}}
        \caption{ Examples of the space-time in which the limit
curve of infinite sequence of closed timelike curves do not
close. }
        \protect 
\end{figure}



\end{document}